\documentclass[aps,prb,twocolumn,showpacs,amsmath,amssymb]{revtex4}

\usepackage{bm}
\usepackage[mathcal]{euscript}
\usepackage{hyperref}
\renewcommand{\vec}[1]{\bm{#1}}
\begin{document}
\preprint{[DRAFT VERSION]}
\title{Comment on ``Magnon wave forms in the presence of a
soliton in two--dimensional antiferromagnets with a staggered field''} %

\author{Denis D. Sheka} %
\email{Denis_Sheka@univ.kiev.ua} %
\affiliation{National Taras Shevchenko University of Kiev, 03127 Kiev, Ukraine} %

\date{March 14 2006}

\begin{abstract}

Very recently Fonseca and Pires [Phys. Rev. B \textbf{73}, 012403
(2006)] have studied the soliton--magnon scattering for the
isotropic antiferromagnet and calculated ``exact'' phase shifts,
which were compared with the ones obtained by the Born
approximation. In this Comment we correct both the soliton and
magnon solutions and point out the way how to study correctly the
scattering problem.

\end{abstract}



\maketitle

The soliton--magnon interaction in 2D magnets is a subject of an
intensive studying more than 10 years. The problem of a magnon
scattering by the Belavin--Polyakov soliton in isotropic magnets, in
particular, antiferromagnets was solves by \citet{Ivanov99}. In a
recent paper \citet{Fonseca06} come back to the soliton--magnon
scattering problem in an isotropic antiferromagnet. The reason is
that authors propose to consider a new type of soliton in the
isotropic antiferromagnet, which is characterized by its internal
precession. In the paper \cite{Fonseca06} authors consider also the
influence of a staggered magnetic field.

In order to describe the soliton structure, the angular
parametrization of the sublattice magnetization vector is involved,
$\vec{n}=\{\sin\theta\cos\phi, \sin\theta \sin\phi, \cos\theta \}$.
The soliton structure is described by the singular distribution of
the  $\phi$--field, $\phi_s = q\varphi-\Omega t$ and the regular one
for the $\theta$--field:
\begin{subequations} \label{eq:theta}
\begin{align} \label{eq:theta-ODE}
&\frac{\mathrm{d}^2\theta_s}{\mathrm{d} r^2} + \frac{1}{r}
\frac{\mathrm{d}\theta_s}{\mathrm{d} r} + \left( k_0^2 -
\frac{q^2}{r^2}\right) \sin\theta_s\cos\theta_s
= h\sin\theta_s,\\
\label{eq:theta-BC} %
&\theta_s(0) = \theta_s(\infty) = 0.
\end{align}
\end{subequations}
Here $r$ and $\varphi$ are the polar coordinates in the XY plane,
$k_0=\Omega/c$. Multiplying Eq.~\eqref{eq:theta-ODE} by
$r^2\mathrm{d}\theta_s/\mathrm{d}r$ and integrating over all $r$
with account of boundary conditions \eqref{eq:theta-BC}, one can
easily obtain the identity \cite{Ivanov87}
\begin{equation} \label{eq:identity}
k_0^2\int_0^\infty \sin^2\theta_s(r)r\mathrm{d}r = h \int_0^\infty
\left[ 1-\cos\theta_s(r) \right] r\mathrm{d}r.
\end{equation}
Note that the identity \eqref{eq:identity} can be satisfied only for
$\theta_s(r)\equiv 0$ in the case of $h\leq0$. However namely this
case, $h\leq0$, corresponds to the ground state $\theta_0=0$: it
minimizes the energy functional (3) of the paper by
\citet{Fonseca06}. The differential problem \eqref{eq:theta} also
has only the trivial solution for $h>0$, this results from the phase
plane analysis.

Thus one can conclude that the differential problem \eqref{eq:theta}
has \emph{only} the trivial solution $\theta_s(r)=0$; hence it has
no sense to consider some nontrivial distribution in the
$\phi$--field, because the soliton does not exist.

Besides the localized soliton solution \citet{Fonseca06} mention
also nolocalized vortex--like solutions. In principle, it is
possible to discuss such solutions when $h>0$, because the ground
state becomes $\theta_0=\arccos H$, see Eqs.~(9) of the paper by
\citet{Fonseca06}. However the energy of such solution does not have
a logarithmic divergence like (12) and (13): the correct form is
mainly determined by the term
\begin{equation*}
\frac{J}{2c^2}\int \mathrm{d}^2 x \sin^2\theta \left(\frac{\partial
\phi}{\partial t}\right)^2 \propto R^2
\end{equation*}
and diverges as the system area, so the precessional vortex solution
also is not preferable.

For the soliton--magnon scattering problem authors come back to the
localized solution. In order to consider magnons in a presence of
the soliton, the following ansatz is involved: $\theta(\vec{r},t) =
\theta_s(\vec{r}) + \eta(\vec{r},t)$. Here authors neglect the
out--of--plane soliton structure, $\theta_s=0$, which corresponds to
our conclusion about the absence of the soliton solution. Thus,
magnons are considered on the following background:
\begin{equation} \label{eq:theta_s}
\theta_s(r) = 0, \qquad \phi_s(\varphi,t) = q\varphi - \Omega t.
\end{equation}
After linearization of Eq.~(4) of the paper authors obtain Eq.~(14),
which leads to the scattering picture.

The reasonable question is how the ``soliton'' \eqref{eq:theta_s},
which is simply a ground state, can scatter magnons? Authors chose
the plane--wave solution of the form $\eta(\vec{r},t) =
\exp(i\vec{k}\cdot\vec{r}-i\omega t)$, which is not a correct
mathematical object, because the real scalar $\eta$ can not be
identified with the complex exponent. The correct form is the real
quantity $\eta(\vec{r},t) = A \cos(\vec{k}\cdot\vec{r}-\omega t)$,
$\phi=\text{const}$, which describes the linearly polarized spin
wave. However this linearly polarized wave is not compatible with
the solution \eqref{eq:theta_s}: after the substitution into Eq.(5)
of Ref.~\onlinecite{Fonseca06}, one can obtain the following
equation
\begin{equation*} \label{eq:14-2}
\frac{q}{r^2}\frac{\partial \eta}{\partial \varphi} =
-\frac{\Omega}{c^2}\frac{\partial \eta}{\partial t},
\end{equation*}
which can not be solved together with Eq.~(14), but authors do not
take it into account. This cause also the wrong dispersion law
(15a).

The correct way is to consider the circular polarized spin wave of
the form $\theta=\text{const}\ll1$, $\phi=\vec{k}\cdot\vec{r} -
\omega t$, which has the same symmetry as a ``soliton'' solution
\eqref{eq:theta_s}. After that the magnons on the soliton background
are described by the linear corrections both to $\theta$ and $\phi$
components and the magnon solution on the background
\eqref{eq:theta_s} has the form similar to Eq.~(17) for both
corrections. However instead of nonanalytic dependence
$\mu=\sqrt{n^2+q^2}$ , the correct index of the Bessel function has
a form $\mu=|n+q|$, see e.g. Ref.~\onlinecite{Ivanov99}. In the case
of the solution \eqref{eq:theta_s} the role of the $q$--term is the
redefinition of the azimuthal quantum numbers $n$. Therefore
the``exact solution'' (17) of the scattering problem should be
reexamined.

We want to stress also that the Born approximation is \emph{not}
adequate for the soliton--magnon scattering problem, see a
discussion in Ref.~\onlinecite{Sheka01}.

This work was supported by the Alexander von Humboldt Foundation.

\end{document}